\documentclass[12pt,twocolumn,preprint, tighten,trackchanges]{aastex63}
\usepackage{natbib}
\usepackage{multirow,color}
\usepackage{bbding}
\usepackage{amssymb}
\usepackage{ulem}
\usepackage{color}
\usepackage{longtable}
\usepackage{array}
\usepackage{bm}
\usepackage{url}
\usepackage{hyperref}
\usepackage{makecell}

\def\gsim{\;\lower4pt\hbox{${\buildrel\displaystyle >\over\sim}$}\;}
\def\lsim{\;\lower4pt\hbox{${\buildrel\displaystyle <\over\sim}$}\;}
\def\grls{\;\lower4pt\hbox{${\buildrel\displaystyle >\over <}$}\;}

\definecolor{darkcyan}{RGB}{0,100,230}

\newcommand{\q}{\color{red}\bf}

\bibpunct{(}{)}{,}{o}{}{,} 

\begin{document}

\title{On the {nature} of photospheric horizontal magnetic field increase in major solar flares} 

\author{Lijuan Liu}
\affiliation{Planetary Environmental and Astrobiological Research Laboratory (PEARL), School of Atmospheric Sciences, Sun Yat-sen University, Zhuhai, Guangdong, 519082, China}
\affiliation{CAS center for Excellence in Comparative Planetology, China}
\affiliation{Key Laboratory of Tropical Atmosphere-Ocean System, Sun Yat-sen University, Ministry of Education, Zhuhai, China}
{\correspondingauthor{Lijuan Liu}}
{\email{liulj8@mail.sysu.edu.cn}}

\author{Zhenjun Zhou}
\affiliation{Planetary Environmental and Astrobiological Research Laboratory (PEARL), School of Atmospheric Sciences, Sun Yat-sen University, Zhuhai, Guangdong, 519082, China}
\affiliation{CAS center for Excellence in Comparative Planetology, China}

\author{Yuming Wang}
\affiliation{CAS Key Laboratory of Geospace Environment, Department of Geophysics and Planetary Sciences, University of Science and Technology of China, Hefei, Anhui, 230026, China}
\affiliation{CAS center for Excellence in Comparative Planetology, China}

\author{Xudong Sun}
\affiliation{Institute for Astronomy, University of Hawai`i at Manoa, Pukalani, HI 96768, USA}

\author{Guoqiang Wang}
\affiliation{Institute of Space Science and Applied Technology, Harbin Institute of Technology, Shenzhen, China}

\begin{abstract}

Rapid increase of horizontal magnetic field ($B_h$) around the flaring polarity inversion line is the most prominent photospheric field change during flares.  
It is considered to be caused by the contraction of flare loops, the {details} behind which is still not fully understood.  
Here we investigate the 
$B_h$-increase in 35 major flares using HMI high-cadence vector magnetograms. 
{We find that $B_h$-increase is always accompanied by the increase of field inclination. It usually initiates near the flare ribbons, 
showing step-like change in between the ribbons.  
In particular, its evolution in early flare phase shows close spatio-temporal correlation to flare ribbons. 
We further find that $B_h$-increase tends to have similar intensity in confined and eruptive flares, but larger spatial-extent in eruptive flares in a statistical sense. 
Its intensity and timescale have inverse and positive correlations to the initial ribbon separations, respectively. 
The results altogether are well consistent with a recent proposed scenario which suggests that the reconnection-driven contraction of flare loops enhances photospheric $B_h$ according to the ideal induction equation, 
providing statistical evidence to the reconnection-driven origin for $B_h$-increase for the first time. }

\end{abstract}

\section{Introduction}\label{sec:intro}

Solar flares are known as the process of sudden energy release caused by magnetic reconnection~\citep{Priest_2002a}. 
They occur on time scale as short as minutes, during which the coronal magnetic field reorganizes rapidly, 
involving the eruption of magnetic flux rope above and the formation of post-flare loops below the reconnecting current sheet {as suggested by the standard flare model~\citep[CHSKP model, see review in][]{Shibata_2011}.} 
The process, which occurs rapidly in tenuous corona, is not expected to exert considerable influence on photosphere since the latter is much denser~\citep{Aulanier_2016}.  
However, observations have revealed a counter-intuitive fact that the photospheric magnetic field does undergo appreciable changes during flares~\citep[see review in][]{Wanghm_2015, Toriumi_2019}. 
The most prominent change is the abrupt, permanent increase of horizontal magnetic field ($B_h$) around the flaring polarity inversion line (PIL), often accompanied by $B_h$ decrease in the peripheral sunspots~\citep[e.g.,][]{Wang_2002a, Sudol_2005, Wangs_2012b, Sun_2017}. 
Different from $B_h$, 
the vertical magnetic field ($B_z$) varies much less without clear pattern~\citep{Sun_2017}. 
Accompanied by the field changes, other photospheric signatures, such as darkening of near-PIL penumbrae and weakening of peripheral penumbrae~\citep{Liu_2005a, Wang_2013h}, sunspot rotation~\citep{Wangs_2014}, increase of field shear 
and inclination 
\citep{Wangs_2012b}, etc., are also found.

The observations indicate an instant feedback from coronal eruptions to the photosphere, 
the {nature} of which is still not fully understood. $B_h$-increase around the flaring PIL, accompanied by the increase of field inclination,   
can be naturally interpreted by the ``tilt'' or contraction of flare loops above the core region. 
This is supported by the topological analysis of coronal magnetic field during flares, 
which reveals the field across PIL collapses toward photosphere after flares~\citep{Li_2011, Sun_2012b, Liuc_2012c}. 
But what exactly causes the field collapse? 
{``Magnetic implosion'' conjecture is often cited as an explanation~\citep{Hudsons_2000}. 
It suggests that the coronal loops must contract to compensate for energy decrease in eruption region according to the rule of energy conservation.} 
The sudden change in corona may also excite an MHD wave which propagates downward and partially penetrates the photosphere to distort the field there~\citep{Hudson_2008, Fletcher_2008, Wheatland_2018}. 
Moreover, 
the photospheric field change is suggested useful for estimating the total Lorentz force change during eruptions, 
therefore useful for estimating the force impulse associated with the coronal mass ejection (CME) momentum~\citep{Fisher_2012, Hudson_2012}.

The details about how flare loops contraction affects the photospheric field are not entirely clear. 
The implosion model mainly predicts the contraction of coronal loops, which seems not guarantee the field near photosphere to respond in a similar way~\citep{Sun_2017}. 
Moreover, although implosion is often related to observed in-eruption contraction of non-erupting coronal loops~\citep[e.g.,][]{Gosain_2012, Simoes_2013, Wang_2018b}, its relation to observed flare loops contraction~\citep{Ji_2007, Liu_2013c} needs clarification.  
Besides, if the flare loops contraction is just morphological, 
it may only affect $B_h$ at the loops footpoints, 
forming disconnected, ribbon-like regions of $B_h$-increase alongside the PIL, 
while the observed $B_h$-increase occurs in a whole region covering the PIL. 
Further detailed explanation is needed. 
Recently, 
\citet{Barczynski_2019} analyzed the 3D magnetic field in a generic MHD simulation of an eruptive flare to draw more details of above process. 
They concluded that the photospheric $B_h$-increase is resulted from the reconnection-driven contraction of flare loops, 
which can be well explained by the ideal induction equation ($\displaystyle \frac{\partial B_{x,y}}{\partial t}{=}{-}B_{x,y}\frac{\partial u_z}{\partial z}$; $u_z$ is the vertical plasma velocity): 
the newly-reconnected flare loops which possess strong line curvature and thus strong magnetic tension at their apex, 
may contract downward and drive reconnection jet. 
The contraction process will brake when approaching the photosphere, 
resulting in negative $\displaystyle \frac{\partial u_z}{\partial z}$ which enhances $B_h$ under the loops through induction equation. 
The work stressed the role of flare-reconnection in enhancing $B_h$. 
Similar conclusions are drawn by a few observations, e.g., 
\citet{Liu_2018g} found a co-temporal and co-spatial evolution trend between $B_h$-increase region and flare ribbons; 
\citet{Wang_2021} discovered that the photospheric field inclination and running penumbral waves were affected by coronal reconnection. 
In addition to these case studies, extensive statistical research is needed.

To explore more details of photospheric field changes in flares, especially the role that flare-reconnection plays, 
we perform a statistical research on the most prominent change 
- $B_h$-increase - in 35 major flares using the high-cadence magnetograms provided by Helioseismic and Magnetic Imager~\citep[HMI;][]{Hoeksema_etal_2014, Sun_2017} onboard {\it Solar Dynamics Observatory (SDO)}. We mainly compared the characteristics of $B_h$-increase and flare ribbons.

\section{Data}\label{sec:data}

HMI measures the Stokes parameters at six wavelengths along Fe {\scriptsize I} 6173 \AA~absorption line,  
based on which the photopsheric vector magnetic field is derived by the Very Fast Inversion of the Stokes Vector algorithm~\citep{Hoeksema_etal_2014}.   
The data has a plate scale of 0\farcs 5, 
with a cadence of 720-s for the regular product, and of 90-s or 135-s for the high-cadence product.  
We select 35 major flares (all larger than M6.0-class, see Table~\ref{tb:props}), including 9 confined and 26 eruptive cases, from the first release of high-cadence data~\citep{Sun_2017}.  
For each flare, we create a set of cutout-maps from full-disk magnetograms to track its source active region (AR). 
The cutout-maps are re-projected from the native Helioprojective-Cartesian coordinate to a local Cartesian cylindrical-equal-area coordinate for easier handling~\citep{Sun_2013a}. 
The formal uncertainty in the spectral line inversion is propagated as error to the data.

We check the flares' CME association through inspecting the {\it SOHO} LASCO CME catalog\footnote{\url{https://cdaw.gsfc.nasa.gov/CME_list/index.html}}, 
and analyze the flare ribbons properties using the 1600~\AA~images provided by Atmospheric Imaging Assembly~\citep[AIA;][]{Lemen_etal_2012} onboard {\it SDO}. 
The data has a plate scale of 0\farcs 6 and a cadence of 24-s. 
The saturation that sometimes appears in AIA images in major flares is corrected by replacing each problematic pixel with a value linearly interpolated from the pixel’s two unsaturated values before and after the saturation period~\citep{Kazachenko_2017}.

\section{Results}\label{sec:res} 

\subsection{An example of $B_h$-increase}\label{subsec:exa}

\subsubsection{Methods}\label{subsec:exa.met}

\begin{figure*}
\begin{center}
\begin{interactive}{animation}{movie_mask.mp4}
\includegraphics[width=0.72\hsize]{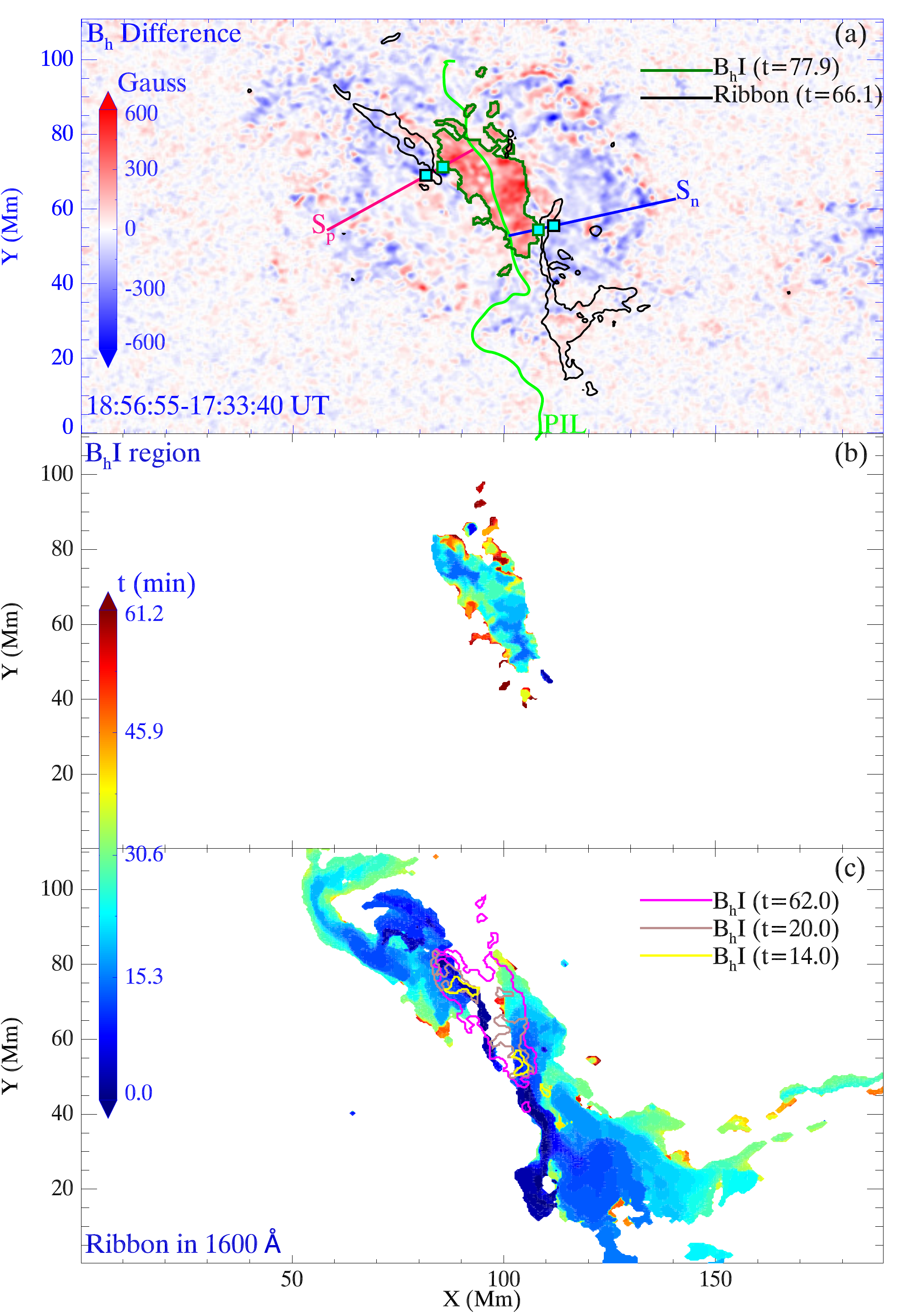} 
\end{interactive}
\caption{
Evolution of $B_h$-increase (abbreviated as ``$B_h$I'' in figures) in case 33. (a) A post-flare $B_h$ difference image. The {blackgreen} contours outline the post-flare mask of $B_h$-increase (pixel value$\geq$120~Gauss); the black contours outline the flare ribbons at one moment for comparison. ``t'' (in minutes) indicates time after {flare start ({\it GOES} start time)}. 
$S_p$ and $S_n$ are slices in positive and negative polarities used to determine the representative points of region fronts ({cyan} squares). 
(b)-(c) Spatio-temporal evolution masks for $B_h$-increase and flare ribbons, with color coded by the time elapsed from {flare start.} 
Colored contours in (c) outline $B_h$-increase regions at different times. 
{The associated animation lasts from 2017-09-06T17:38 to 2017-09-06T18:56, showing the $B_h$-increase and ribbons evolution. 
Its frames have similar layout as the figure.}  
}\label{fig:exa-mask} 
\end{center}
\end{figure*}

\begin{figure*}
\begin{center}
\begin{interactive}{animation}{movie_para.mp4}
\includegraphics[width=0.68\hsize]{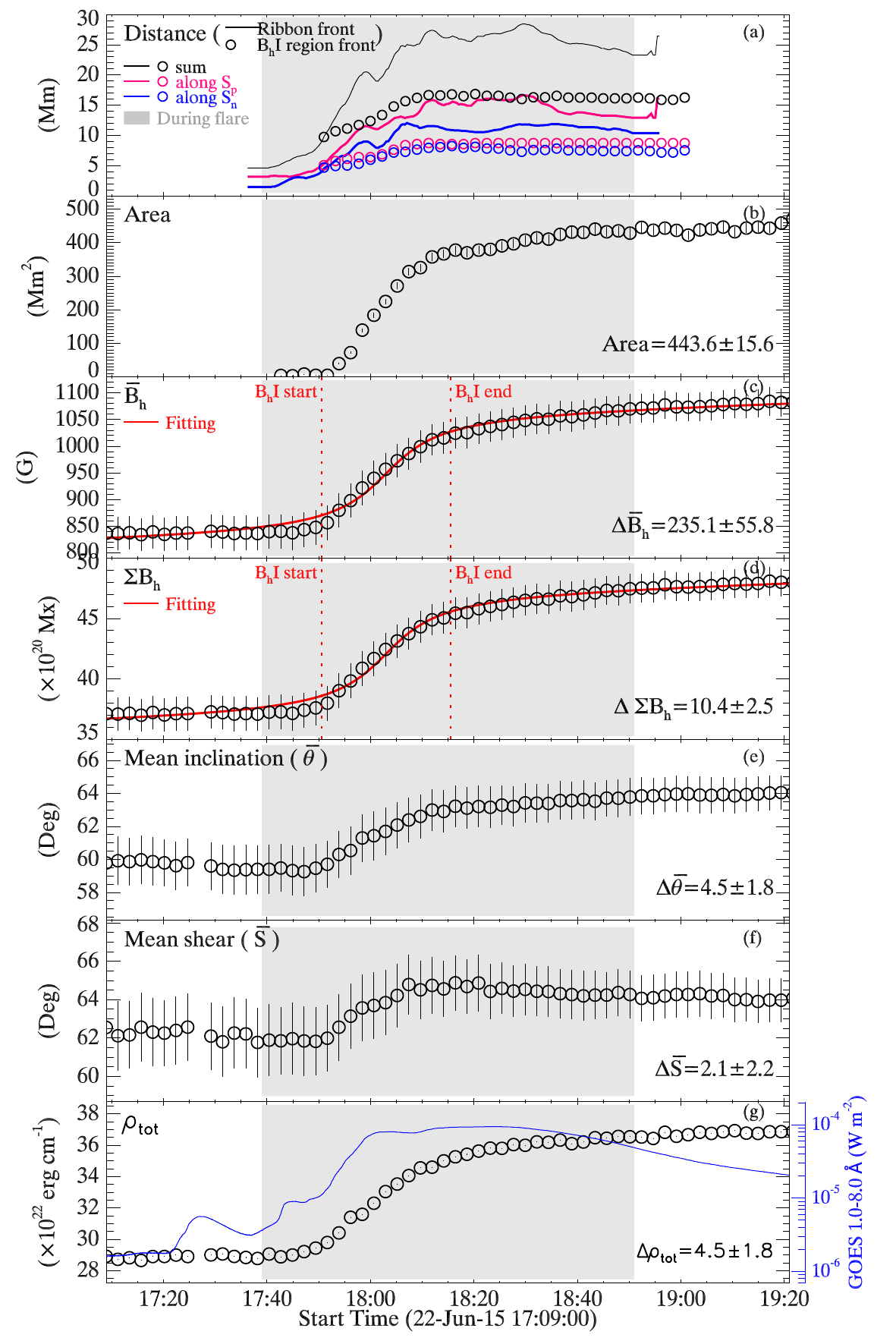} 
\end{interactive}
\caption{(a) Evolution of distances between the PIL and the fronts of  $B_h$-increase region (circles) and flare ribbons (lines) determined by the slices in Figure~\ref{fig:exa-mask}(a). 
Distances in both polarities (blue and red) and as sum (black) are shown. (b)-(g) Parameters quantifying $B_h$-increase. The vertical-dashed lines in (c) and (d) mark the start and end timings of step-like changes obtained from fitting. 
{The associated animation lasts from 2017-09-06T17:38 to 2017-09-06T18:56, showing parameters evolution accompanied by $B_h$-increase region and ribbons evolution. Each frame contains parameters panels (similar as the figure) and two insets of $B_h$-increase region and flare ribbons (similar as Figure~\ref{fig:exa-mask}(b)-(c)). The frame time is indicated by a vertical line in parameters panels.} 
}\label{fig:exa-para} 
\end{center}
\end{figure*}

We analyze the evolution of flare-related $B_h$-increase case-by-case. 
An M6.6-class flare ({SOL2015-06-22T17:39}, case 33 in Table~\ref{tb:props}) is shown as an example {(Figure~\ref{fig:exa-mask} and Figure~\ref{fig:exa-para}).} 
We firstly scrutinize the 
vector magnetograms and construct two masks of $B_h$-increase, 
including a post-flare mask used to pinpoint the final $B_h$-increase region 
(Figure~\ref{fig:exa-mask}(a)), 
and a spatio-temporal mask used to check the propagation of $B_h$-increase (Figure~\ref{fig:exa-mask}(b)). 
The first mask consists of pixels where post-flare $B_h$-increase exceeds 120~Gauss (red patches enclosed by {darkgreen} 
contours in Figure~\ref{fig:exa-mask}(a)). The pixels are identified from a $B_h$ difference image 
constructed by subtracting a pre-flare $B_h$ map from a post-flare $B_h$ map. 
The threshold 
for identifying $B_h$-increase, 120~Gauss, is slightly higher than HMI $B_h$ uncertainty~\citep{Hoeksema_etal_2014}. 
The second mask records the time when $B_h$-increase firstly reaches 120~Gauss in each pixel.   
A similar method is performed on AIA 1600~\AA~images to obtain the spatio-temporal mask for flare ribbons (Figure~\ref{fig:exa-mask}(c)). 
We use a threshold 
several times larger than the median pixel value of all 1600~\AA~maps to identify the ribbon pixels~\citep{Kazachenko_2017}.

We further track the fronts of $B_h$-increase region and flare ribbons to compare their propagation details. 
In each polarity, we select a representative point from the region front (either of $B_h$-increase region or of flare ribbons), and track the point through measuring its distance to the PIL.  
The representative point  ({cyan} squares in Figure~\ref{fig:exa-mask}(a)) is determined as the 
intersection between region boundary and a slice  perpendicular to the source PIL ($S_p$ and $S_n$ in Figure~\ref{fig:exa-mask}(a)). 
We choose the slices as close as possible to ribbons initiation positions. 
The results are shown in {Figure~\ref{fig:exa-para}(a).}

We also calculate a few parameters from the $B_h$-increase region to quantify its evolution, 
including the area, mean $B_h$ ($\overline{B}_h$), total $B_h$ ($\Sigma B_h$), mean shear angle ($\overline{S}$), mean inclination angle ($\overline{\theta}$; with respect to solar normal), and the {proxy of total photospheric excess magnetic energy density~\citep[$\rho_{tot}$,][]{Wang_1996, Leka_Barnes_2003a}. 
Calculation details are shown in Table~\ref{tb:paras}}. 
$\overline{B}_h$ quantifies the intensity, while $\Sigma B_h$ and area measure the extension, of $B_h$-increase. 
$\overline{\theta}$ measures the field inclination.  
{$\overline{S}$ and $\rho_{tot}$ quantify the deviation of magnetic field from its potential state to some extent.} 
If applicable, the temporal evolution of a parameter $P$ is further fitted by a step-like function 
\begin{equation}\label{eq:step}
\displaystyle P(t){=}a{+}bt{+}c{\left\{1{+}\frac{2}{\pi} arctan[n(t-t_m)]\right\}}, 
\end{equation} 
where $a$, $b$, $c$, $n$, and $t_m$ are free parameters~\citep{Sudol_2005, Sun_2017}. The term $a{+}bt$ indicates the linear evolution 
in addition to step-like change; the parameter change is calculated by $\Delta P{=}2c$; the change timescale is given by $\tau{=}\pi/n$; $t_m$ is the change mid-time. 
We use IDL procedure {\it mpfit.pro}\footnote{\url{https://pages.physics.wisc.edu/~craigm/idl/fitting.html}} to perform the fitting. 
The parameters evolution, with errors propagated from the formal uncertainty, are shown in {Figure~\ref{fig:exa-para}(b)-(g). 
The {\it GOES} 1-8~\AA~flux is shown for comparison (blue curve in Figure~\ref{fig:exa-para}(g)).}

\subsubsection{Features of $B_h$-increase} \label{subsec:exa.res} 

The high-cadence data reveals the evolution details of $B_h$-increase. 
It is seen that $B_h$-increase starts from two kernels alongside the PIL (darkblue patches in Figure~\ref{fig:exa-mask}(b) and yellow contours in Figure~\ref{fig:exa-mask}(c)), roughly coinciding with the flare ribbons initial positions. 
It then extends out (see Figure~\ref{fig:exa-mask}-associated movie), filling a whole region covering part PIL. 
Evolution of the regions fronts {(Figure~\ref{fig:exa-para}(a))}  
further reveals that $B_h$-increase does follow the ribbons propagation in early flare phase,  
but remains almost still afterwards while the ribbons propagate further.

For the parameters quantifying $B_h$-increase {(Figure~\ref{fig:exa-para}(b)-(g))},  
the area reaches around 443.6~Mm$^2$ after flare. 
$\overline{B}_h$ and $\Sigma B_h$ both show step-like increase which can be fitted by Equation~\ref{eq:step}. 
The fitting reveals that $\overline{B}_h$ begins to increase 11.6 minutes after the {flare start ({\it GOES} start time)}, 
and lasts for 24.9 minutes.  
Its post-flare change ($\Delta \overline{B}_h$) is $235.1$ Gauss. 
The fitting of $\Sigma B_h$ yields similar results. 
Its post-flare change ($\Delta \Sigma B_h$) is $10.4\times 10^{20}$~Mx. 
$\overline{\theta}$, $\overline{s}$, and $\rho_{tot}$ also show step-like change here, but their evolution in most other cases are too complicated to fit. To avoid the bias, we calculate the three parameters' change here, as well as in other cases, by subtracting the pre-flare value from the post-flare value instead of fitting {(see 
details in Table~\ref{tb:paras})}. Here their changes ($\Delta \overline{\theta}$, $\Delta \overline{S}$, and $\Delta \rho_{tot}$) are $4.5^{\circ}$, $2.1^{\circ}$, and $4.5 \times 10^{22} {\rm erg\ cm^{-1}}$, respectively, indicating the field in $B_h$-increase region becomes more {inclined and sheared} after the flare.

\subsection{Statistics of $B_h$-increase}\label{subsec:sta}

\begin{figure*}
\begin{center}
\epsscale{1.18}
\plotone{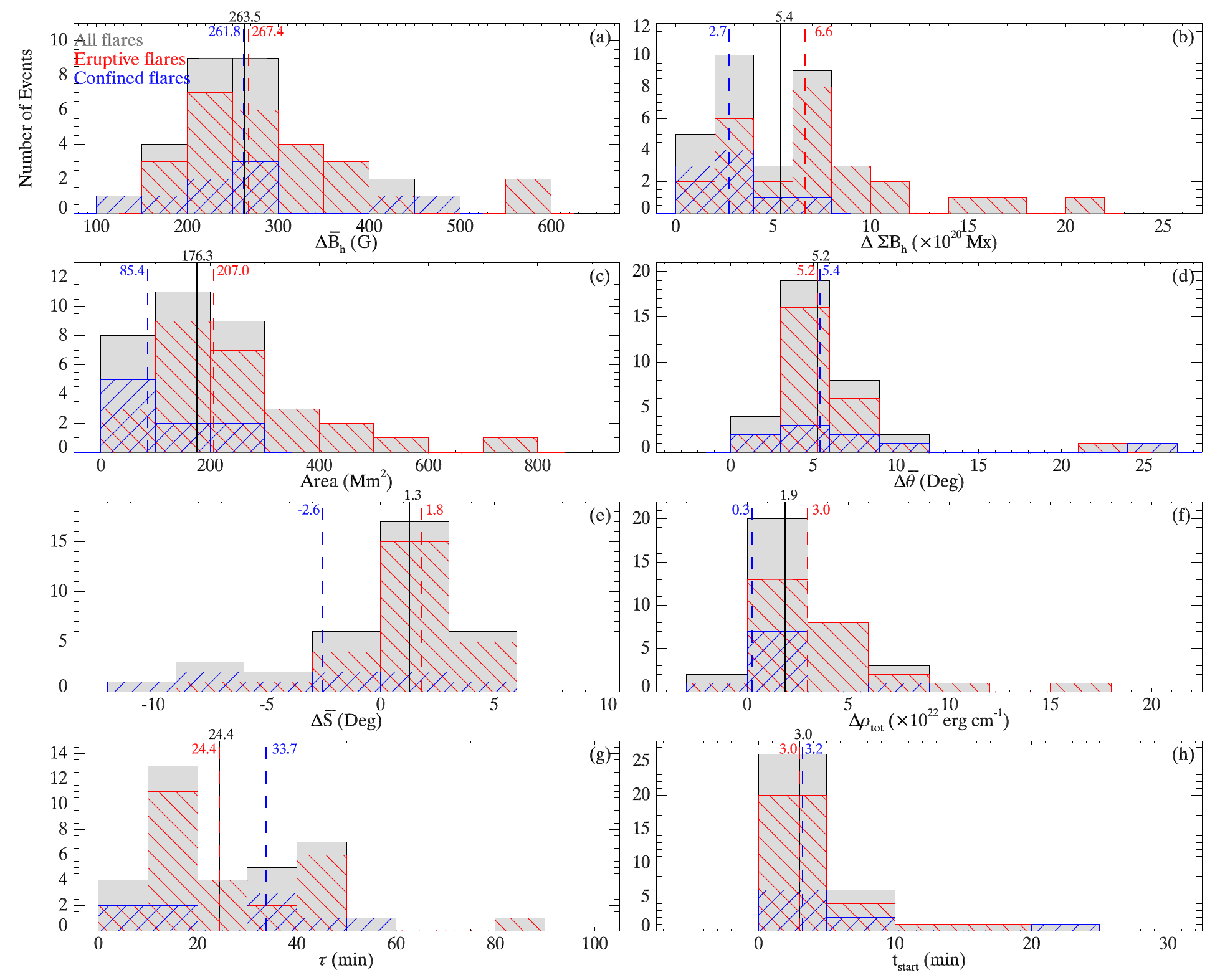} 
\caption{Distributions of the post-flare change of $B_h$-increase parameters.   
In each panel, the vertical lines, with the digits above, 
mark the parameters medians for corresponding samples (black, red and blue for all, eruptive and confined flares, respectively). 
}\label{fig:dist} 
\end{center}
\end{figure*}

Using the above methods, we identify $B_h$-increase in all cases 
and perform a statistical research. 
We found the post-flare change of $\overline{B}_h$, $\Delta \overline{B}_h$, ranges from $144.5$ G to $573.4$ G in the sample, 
having a median of $263.5$ G (Figure~\ref{fig:dist}(a)). 
The change of $\Sigma B_h$, $\Delta \Sigma B_h$, has a median of $5.4{\times} 10^{20}$~Mx (Figure~\ref{fig:dist}(b)). 
The area has a median of $176.3$~Mm$^2$ (Figure~\ref{fig:dist}(c)). 
The change of $\overline{\theta}$ ($\Delta \overline{\theta}$) shows increase in all cases as well, having a median of $5.2^{\circ}$ (Figure~\ref{fig:dist}(d)). The change of $\overline{S}$ ($\Delta \overline{S}$) shows increase in about $2/3\ (23)$ cases. 
Its median is $1.3^{\circ}$ (Figure~\ref{fig:dist}(e)). 
The change of $\rho_{tot}$ ($\Delta \rho_{tot}$) displays increase in $94\%\ (33)$ cases, having a median of $1.9{\times} 10^{22} {\rm erg\ cm^{-1}}$ (Figure~\ref{fig:dist}(f)). 
The timescale of $B_h$-increase, $\tau$, ranges from 5.8~minutes to 84.4 minutes, having a median of 24.4 minutes (Figure~\ref{fig:dist}(g)). 
The start time of $B_h$-increase, $t_{start}$, is generally small, 
having a median of 3.0 minutes (Figure~\ref{fig:dist}(h)). 
$\tau$ and $t_{start}$ are quoted from the $\overline{B}_h$ evolution fitting.  
The results are consistent with previous observations that the 
field in $B_h$-increase region becomes more inclined and sheared after flares.

When dividing the cases into confined and eruptive groups (Figure~\ref{fig:dist}), 
the {intensive parameter} of $B_h$-increase, 
$\Delta \overline{B}_h$, shows very close medians in the two sub-samples,
suggesting a weak statistical difference. 
By contrast, the extensive parameters, 
$\Delta \Sigma B_h$ and area, 
shows larger medians in eruptive flares. 
Similarly, $\Delta \overline{S}$ and $\Delta \rho_{tot}$, 
also have larger medians in eruptive flares. 
$\Delta \overline{\theta}$  and $t_{start}$ show no significant statistical difference in two sub-samples.  
Only $\tau$ has larger median in confined flares. 
These suggest that $B_h$-increase occurs in either confined or eruptive flares, 
having similar intensity in both kinds 
but larger spatial-extent in eruptive flares.

\begin{figure*}
\begin{center}
\epsscale{1.18}
\plotone{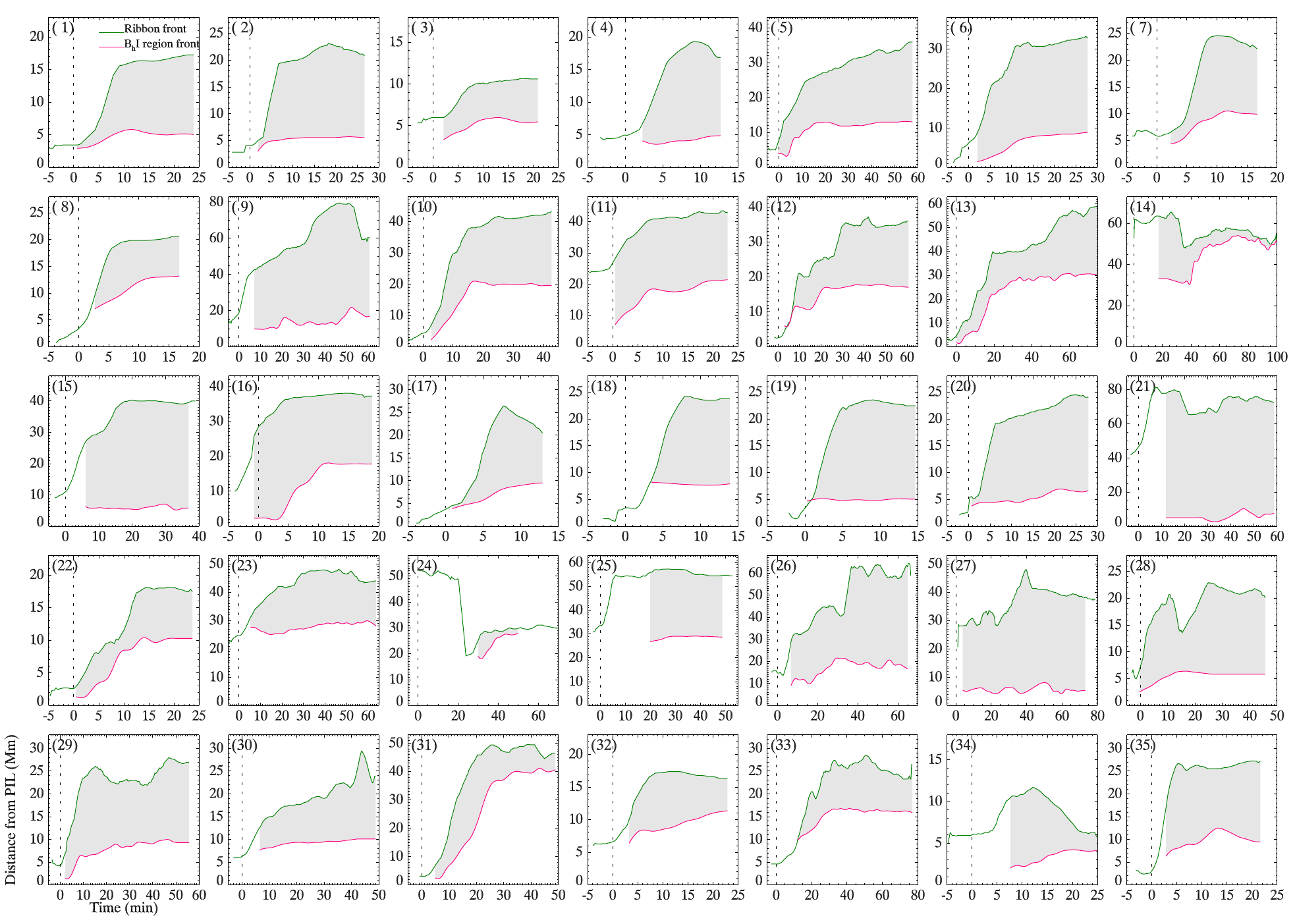}
\caption{Evolution of the distances from the regions fronts to the flaring PILs in all cases, 
with red (green) for $B_h$-increase region (flare ribbons). 
The subtitles indicate case numbers in Table~\ref{tb:props}. 
The vertical lines mark the {flares start.}  
The shaded regions mark the co-existence duration of $B_h$-increase and flare ribbons.}\label{fig:rib} 
\end{center}
\end{figure*}

We also compare the propagation of $B_h$-increase region and flare ribbons in all cases through tracking their fronts (Figure~\ref{fig:rib}). It is seen that $B_h$-increase always appears after the flare ribbons appear, 
with $60\%\ (21)$ cases initiating quite close to the ribbons (with separation $\leq5$~Mm). 
The region then widens, manifested as front progressing, in early phase of most flares.   
Only in 7 cases (e.g., case 4), the region front barely moves after appearance (progressing $\leq 2$~Mm). 
Throughout the flare, the $B_h$-increase region evolves in between the straight parts of flare ribbons, 
and stops growing before the ribbons stop progressing, 
as suggested by the increasing distance between their fronts. 
The results indicate that in general, $B_h$-increase and flare ribbons have close spatio-temporal 
correlations in early flare phase.

\begin{figure*}
\begin{center}
\epsscale{1.1}
\plotone{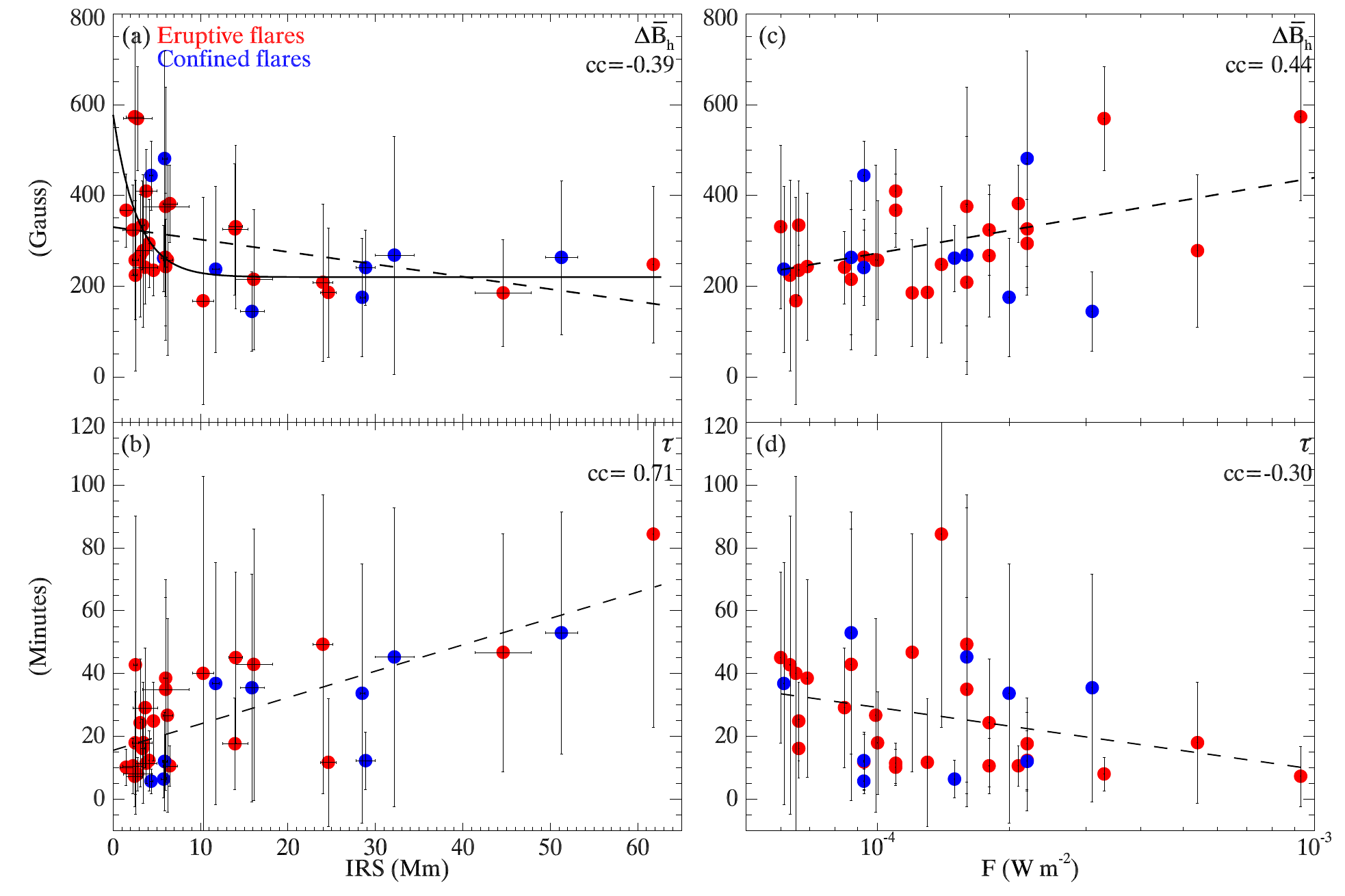}
\caption{(a)-(b) Scatter plots between two $B_h$-increase parameters and initial ribbons separation (IRS).  
The solid curve in (a) represents an exponential fitting. 
The dashed line represents the linear fitting. 
``cc'' indicates the correlation coefficient. 
(c)-(d) Scatter plots between the parameters and flare magnitude shown for comparison. 
}\label{fig:fit} 
\end{center}
\end{figure*}

The summed value of distances from ribbons fronts to the PIL in both positive and negative polarities (green curves in Figure~\ref{fig:rib}) represents the separation between flare ribbons. 
Since the straight parts of flare ribbons highlight the footpoints of flare loops as suggested in the 3D extension of standard flare model~\citep{Aulanier_2012, Janvier_2013}, 
the ribbons separation can indicate the loops height if consider a nearly-semicircular shape of the loops~\citep{Fletcher_2008, Thalmann_etal_2015, Kusano_2020}.  
Therefore the ribbons separation obtained in very early flare stage, 
defined as initial ribbons separation (IRS) here {(see calculation details in Table~\ref{tb:paras}),} 
indicates the height where early flare loops are formed, i.e., where flare reconnection initiates. 
We then check the correlation between the IRS and $B_h$-increase parameters.   
Interestingly, 
we find a rough inverse correlation between $\Delta \overline{B}_h$ and IRS (Figure~\ref{fig:fit}(a)). 
Their relation can be roughly fitted by an exponential decaying function 
in the form of $\Delta \overline{B}_h{=}220.0{+}357.6e^{-0.36{\times}IRS}$ (reduced chi-square $\chi_r^2{=}0.61$). 
Given the large uncertainty of $\Delta \overline{B}_h$, it is more appropriate to see the fitting only as a reference.  
Moreover, the timescale $\tau$ of $B_h$-increase shows a significant correlation with IRS (cc=0.71; Figure~\ref{fig:fit}(b)). 
These indicate that smaller initial ribbon separation tends to be accompanied by $B_h$-increase of larger intensity and shorter timescale.

We also check the correlation between flare magnitude and the parameters, 
and find an overall weak trend that larger flares are accompanied by $B_h$-increase of larger intensity and shorter timescale (Figure~\ref{fig:fit}(c)-(d)).

\section{Summary and Discussion}\label{sec:sum}

In this work, we investigate the photospheric horizontal field increase in 35 major flares using the HMI high-cadence field data, and obtain the following results which may deepen our understanding to flare-related $B_h$-increase: 

\begin{itemize}

\item[1)]
  
$B_h$-increase appears in every case,  
accompanied by the field inclination increase. 
This supports that the magnetic field rooted in this region becomes more inclined after the flare.   

\item[2)]

$B_h$-increase usually initiates near the flare ribbons, showing step-like change which evolves in a limited region in between the straight parts of flare ribbons.   
{In particular, its evolution in early flare phase shows close spatio-temporal correlation to flare ribbons.} 
Since the ribbons highlight loops footpoints, 
these suggest that $B_h$-increase is closely related to flare loops evolution. 

\item[3)]

The {intensive parameter} of $B_h$-increase (mean $B_h$) shows no statistical difference in confined and eruptive flares, while the extensive parameters (total $B_h$ and area) tend to be larger in eruptive flares. 
These suggest that the {process leading to $B_h$-increase} 
tends to affect larger photospheric area during CME-associated flares.

\item[4)] 
The intensity and timescale of $B_h$-increase are inversely and positively correlated to the initial ribbons separation, respectively. 
Since initial ribbon separation indicates the height where flare-reconnection initiates (see 
Section~\ref{subsec:sta}), 
these suggest that the process leading to $B_h$-increase 
may start along with the flare-reconnection, and then propagates downward, resulting in faster and stronger photospheric effect 
when the propagating path is shorter (lower reconnection height).

\end{itemize}


The above results altogether support that $B_h$-increase is resulted from the flare reconnection-driven contraction of flare loops,     
governed by induction equation~\citep{Barczynski_2019}. 
{The first finding (item 1)} supports the occurrence of flare loops contraction, consistent with previous studies (see 
Section~\ref{sec:intro}). 
{The second} further supports a flare loops-related origin for $B_h$-increase. 
{The third} has a natural interpretation from the 
standard flare model: the upward motion of CMEs in eruptive flares tends to 
result in more flare loops which may sweep wider photospheric area. 
{The fourth} is well explained when considering the induction equation: smaller initial ribbon separation indicates lower, shorter flare loops formed in early flare phase, which may relax downward faster and brake sharply with the tension-driven jet,  
resulting in larger $\displaystyle \frac{\partial u_z}{\partial z}$ which amplifies photospheric $B_h$ more quickly and significantly.  
The higher, longer loops formed later tend to behave oppositely, 
explaining why $B_h$-increase only follows the ribbons 
in early flare phase. 
This is analogous to the phenomenon revealed in simulation: $B_h$-increase occurs slower in regions later swept by current ribbons~\citep{Barczynski_2019}. 
{Note the latter three findings are systematically identified here for the first time.} 
Besides above results, the identified shear increase is also consistent with the shear transfer process suggested in 3D flare model.

{
Our results suggest that $B_h$-increase is consistent with the standard flare model. Although the model is proposed for eruptive flares, the difference of $B_h$-increase in confined and eruptive flares identified here is not striking (see item 3 above). 
This might because even some confined flares may still involve confined flux rope eruptions. 
A flare showing observable signatures of confined eruption of possible flux ropes  (e.g., filaments or hot channels) is called failed eruption~\citep{Jihs_2003}. 
In that case, the flux rope also rises, leading to the formation of current sheet below, but fails to propagate out due to strong confinement above~\citep{Torok_Kliem_2005}. 
The flare thus tends to have similar properties to an eruptive flare~\citep{Harra_2016}. 
However, its ribbons may sweep less large photospheric area as the flux rope stops ascending earlier, 
consistent with our findings above (item 3). 
For the confined flare not involving a flux rope, 
observations suggest that the flare-reconnection may occur between multiple magnetic systems, and about one third of observed confined flares belongs to this category~\citep{Li_2019}.  
Considering that we have around $27\%$ (9/35) confined flares, 
the portion of the category is quite small, which may not significantly affect our statistical results despite what their $B_h$-increase properties are.} 

{The implosion conjecture~\citep{Hudsons_2000} is often referred to explain $B_h$-increase (see 
Section~\ref{sec:intro}). Although our results support a reconnection-driven origin for $B_h$-increase, 
the implosion should still be consistent with the 
process since it is just a restatement of the universal rule of energy conservation.} 
The conservation of momentum during coronal eruptions~\citep{Hudson_2012, Fisher_2012} is also referred to explain the flare-related $B_h$-increase sometimes~\citep[e.g.,][]{Barczynski_2019, Wang_2021}. 
But to our understanding, 
the CME momentum, and corresponding downward momentum, is related to the photospheric field change through Lorentz force change under some approximation, and only sets a lower limit to Lorentz force impulse~\citep{Fisher_2012}. 
It may not be appropriate to take the momentum conservation as an independent cause for $B_h$-increase.
The identified relatively larger total $B_h$-increase in eruptive flares indeed suggests a larger Lorentz force change in CMEs.

To summarize, our results support that $B_h$-increase during flares is very likely to be caused by the reconnection-driven contraction of flare loops, and is governed by induction equation. 


\clearpage

\setlength{\tabcolsep}{0.1mm}{

\begin{longrotatetable}
\startlongtable 

\begin{deluxetable*}{c|ccccc|ccccccccc} 

\tabletypesize{\footnotesize} 

\tablecaption{Flare information and $B_h$-increase characteristics.}\label{tb:props} 
\tablehead{
\colhead{No.} & \multicolumn{5}{c}{Flare} & \multicolumn{9}{c}{Parameters} 
}
\startdata 
    & Class\phantom{aa}  & Start\phantom{aa} & Location\phantom{a} & AR  & CME$^{\hyperlink{tb1:a1}{a}}$ & $\Delta \overline{B}_h$ & $\Delta \Sigma B_h$ & Area & $\Delta \overline{\theta}$ & $\Delta \overline{S}$ & $\Delta \rho_{total}$ & $\tau$ & $t_{start}$ & IRS$^{\hyperlink{tb1:b1}{b}}$ \\ 
    & & & & & & Gauss &$\times10^{20}$ Mx & Mm$^2$ & degree & degree  & $\times10^{22}$erg cm$^{-1}$ & mins\phantom{a} & mins & Mm \\
\hline
1 & M6.6 & { SOL2011-02-13T17:28} & S20E05 & 11158 & Y & $334.5{\pm} 98.2$ & $  3.4{\pm}  1.0$ & $101.6{\pm}  3.9$ & $  3.0{\pm}  2.0$ & $  1.0{\pm}  2.1$ & $  2.04{\pm}  0.02$ &  0.6 &16.2 &$  3.4{\pm} 0.2$ \\
2 & X2.2 & { SOL2011-02-15T01:44} & S20W12 & 11158 & Y & $294.2{\pm} 97.3$ & $  7.8{\pm}  2.6$ & $252.5{\pm} 11.4$ & $  6.0{\pm}  2.3$ & $  3.5{\pm}  2.5$ & $  4.66{\pm}  0.04$ &  1.4 &12.4 &$  4.1{\pm} 0.0$ \\
3 & X1.5 & { SOL2011-03-09T23:13} & N08W11 & 11166 & N & $261.8{\pm} 73.2$ & $  1.3{\pm}  0.4$ & $ 52 .3{\pm}  2.7$ & $  2.0{\pm}  2.3$ & $ -4.7{\pm}  2.4$ & $  0.25{\pm}  0.02$ &  2.8 & 6.4 &$  5.8{\pm} 0.3$ \\
4 & M9.3 & { SOL2011-07-30T02:04} & N14E35 & 11261 & N & $444.0{\pm} 75.6$ & $  2.7{\pm}  0.5$ & $ 59.8{\pm}  4.4$ & $  6.3{\pm}  3.7$ & $ -8.2{\pm}  4.2$ & $  1.18{\pm}  0.03$ &  3.2 & 5.8 &$  4.3{\pm} 0.2$ \\
5 & M6.0 & { SOL2011-08-03T13:17} & N16W30 & 11261 & Y & $331.1{\pm} 179.9$ & $  6.9{\pm}  3.8$ & $206.2{\pm} 11.8$ & $  5.7{\pm}  3.3$ & $  2.2{\pm}  4.1$ & $  1.88{\pm}  0.02$ &  1.1 &45.1 &$ 14.0{\pm} 0.8$ \\
6 & M9.3 & { SOL2011-08-04T03:41} & N16W38 & 11261 & Y & $263.7{\pm} 85.1$ & $  6.7{\pm}  2.4$ & $258.5{\pm} 14.7$ & $  4.0{\pm}  3.8$ & $  1.3{\pm}  4.5$ & $  3.64{\pm}  0.03$ &  5.7 &11.9 &$  6.0{\pm} 0.3$ \\
7 & X2.1 & { SOL2011-09-06T22:12} & N14W18 & 11283 & Y & $382.0{\pm} 84.6$ & $  6.6{\pm}  1.6$ & $173.9{\pm}  8.7$ & $  6.3{\pm}  3.3$ & $  2.1{\pm}  4.1$ & $  3.41{\pm}  0.03$ &  3.8 &10.7 &$  6.5{\pm} 0.8$ \\
8 & X1.8 & { SOL2011-09-07T22:32} & N14W31 & 11283 & Y & $324.1{\pm} 98.9$ & $  6.4{\pm}  2.2$ & $185.0{\pm} 14.4$ & $  4.9{\pm}  4.8$ & $ -2.1{\pm}  5.7$ & $  1.75{\pm}  0.03$ &  3.4 &10.6 &$  2.3{\pm} 1.2$ \\
9 & M8.7 & { SOL2012-01-23T03:38} & N33W21 & 11402 & Y & $215.1{\pm} 154.5$ & $  3.6{\pm}  2.7$ & $166.3{\pm} 17.4$ & $  7.1{\pm}  8.8$ & $ -7.8{\pm}  9.7$ & $  1.06{\pm}  0.03$ &  5.3 &42.9 &$ 16.1{\pm} 2.1$ \\
10 & X5.4 & { SOL2012-03-07T00:02} & N18E31 & 11429 & Y & $278.4{\pm} 168.1$ & $ 20.5{\pm} 13.2$ & $720.6{\pm} 51.8$ & $  9.9{\pm}  3.5$ & $  2.4{\pm}  3.7$ & $ 11.02{\pm}  0.08$ &  5.7 &18.0 &$  3.4{\pm} 1.0$ \\
11 & X1.3 & { SOL2012-03-07T01:05} & N15E26 & 11429 & Y & $186.3{\pm} 142.0$ & $  3.4{\pm}  2.8$ & $176.3{\pm} 18.2$ & $  5.3{\pm}  6.0$ & $  2.7{\pm}  6.6$ & $  1.28{\pm}  0.05$ &  0.3 &11.8 &$ 24.6{\pm} 0.9$ \\
12 & M6.3 & { SOL2012-03-09T03:22} & N15W03 & 11429 & Y & $224.2{\pm}210.5$ & $  6.2{\pm}  5.9$ & $275.5{\pm} 18.6$ & $  3.2{\pm}  4.3$ & $ -2.5{\pm}  5.1$ & $  2.13{\pm}  0.03$ &  2.2 &42.8 &$  2.5{\pm} 0.2$ \\
13 & M8.4 & { SOL2012-03-10T17:15} & N17W24 & 11429 & Y & $241.6{\pm} 80.1$ & $  9.2{\pm}  3.1$ & $405.7{\pm} 25.7$ & $  7.2{\pm}  5.2$ & $  3.6{\pm}  6.0$ & $  5.09{\pm}  0.03$ &  0.8 &29.2 &$  3.7{\pm} 1.4$ \\
14 & X1.4 & { SOL2012-07-12T15:37} & S13W03 & 11520 & Y & $247.9{\pm} 172.8$ & $  7.2{\pm}  4.9$ & $292.2{\pm} 18.5$ & $  2.3{\pm}  5.1$ & $ -1.0{\pm}  8.8$ & $  1.62{\pm}  0.02$ & 17.2 &84.4 &$ 61.8{\pm} 0.0$ \\
15 & M6.5 & { SOL2013-04-11T06:55} & N09E13 & 11719 & Y & $167.8{\pm} 228.7$ & $  0.6{\pm}  0.8$ & $ 36.8{\pm}  3.6$ & $  3.8{\pm}  4.6$ & $  2.6{\pm}  7.1$ & $  0.12{\pm}  0.01$ &  0.2 &40.1 &$ 10.3{\pm} 1.2$ \\
16 & M9.3 & { SOL2013-10-24T00:21} & S09E10 & 11877 & N & $241.1{\pm} 83.3$ & $  2.9{\pm}  1.0$ & $118.2{\pm}  6.9$ & $  3.2{\pm}  3.0$ & $  2.9{\pm}  4.5$ & $  1.40{\pm}  0.02$ &  0.1 &12.3 &$ 28.9{\pm} 1.1$ \\
17 & X3.3 & { SOL2013-11-05T22:07} & S12E44 & 11890 & Y & $569.4{\pm} 114.7$ & $  4.8{\pm}  1.1$ & $ 81.5{\pm}  3.7$ & $  5.2{\pm}  2.7$ & $  3.4{\pm}  2.6$ & $  4.83{\pm}  0.05$ &  3.5 & 8.1 &$  2.8{\pm} 1.6$ \\
18 & X1.1 & { SOL2013-11-08T04:20} & S13E15 & 11890 & Y & $409.7{\pm} 92.9$ & $  4.4{\pm}  1.0$ & $107.7{\pm}  4.5$ & $  4.8{\pm}  2.0$ & $  3.7{\pm}  2.1$ & $  3.42{\pm}  0.03$ &  0.6 &11.5 &$  3.8{\pm} 1.2$ \\
19 & X1.1 & { SOL2013-11-10T05:08} & S13W13 & 11890 & Y & $367.4{\pm} 80.5$ & $  3.7{\pm}  0.9$ & $106.0{\pm}  5.1$ & $  5.9{\pm}  3.0$ & $  1.0{\pm}  3.7$ & $  3.00{\pm}  0.03$ &  0.7 &10.2 &$  1.5{\pm} 0.8$ \\
20 & M9.9 & { SOL2014-01-01T18:40} & S16W45 & 11936 & Y & $257.9{\pm}209.1$ & $  3.6{\pm}  3.4$ & $155.2{\pm} 20.8$ & $ 21.4{\pm} 12.2$ & $ -4.4{\pm} 13.1$ & $ -0.29{\pm}  0.02$ &  1.4 &26.7 &$  6.2{\pm} 0.5$ \\
21 & X1.2 & { SOL2014-01-07T18:04} & S12W08 & 11944 & Y & $185.4{\pm} 118.1$ & $  0.8{\pm}  0.5$ & $ 42.5{\pm}  2.7$ & $  1.8{\pm}  2.2$ & $  1.3{\pm}  4.9$ & $  0.01{\pm}  0.00$ &  9.1 &46.8 &$ 44.6{\pm} 3.2$ \\
22 & X1.0 & { SOL2014-03-29T17:35} & N10W32 & 12017 & Y & $257.4{\pm} 130.8$ & $  3.3{\pm}  2.1$ & $138.7{\pm} 12.4$ & $  7.0{\pm}  5.6$ & $ -3.0{\pm}  6.7$ & $  0.19{\pm}  0.02$ &  3.2 &18.0 &$  2.5{\pm} 0.4$ \\
23 & X1.6 & { SOL2014-09-10T17:21} & N11E05 & 12158 & Y & $208.5{\pm} 173.2$ & $  8.0{\pm}  7.0$ & $381.7{\pm} 22.4$ & $  3.4{\pm}  2.6$ & $  0.1{\pm}  4.1$ & $  1.88{\pm}  0.02$ &  1.2 &49.3 &$ 24.0{\pm} 1.1$ \\
24 & M8.7 & { SOL2014-10-22T01:16} & S13E21 & 12192 & N & $263.5{\pm} 169.8$ & $  6.9{\pm}  4.6$ & $261.9{\pm} 13.1$ & $  3.2{\pm}  2.3$ & $ -2.6{\pm}  2.6$ & $  0.71{\pm}  0.02$ &  5.4 &53.0 &$ 51.3{\pm} 1.9$ \\
25 & X1.6 & { SOL2014-10-22T14:02} & S14E13 & 12192 & N & $268.5{\pm} 262.0$ & $  0.5{\pm}  0.6$ & $ 17.4{\pm}  0.8$ & $  0.9{\pm}  2.5$ & $  0.5{\pm}  2.7$ & $  0.25{\pm}  0.01$ &  4.2 &45.3 &$ 32.2{\pm} 2.2$ \\
26 & X3.1 & { SOL2014-10-24T21:07} & S22W21 & 12192 & N & $144.5{\pm} 87.3$ & $  3.6{\pm}  2.2$ & $250.1{\pm} 21.4$ & $  5.4{\pm}  4.5$ & $ -0.0{\pm}  5.9$ & $  0.20{\pm}  0.02$ &  0.1 &35.5 &$ 15.9{\pm} 1.4$ \\
27 & X2.0 & { SOL2014-10-26T10:04} & S14W37 & 12192 & N & $175.5{\pm} 131.0$ & $  1.4{\pm}  0.8$ & $ 58.2{\pm} 10.5$ & $ 25.4{\pm}  9.8$ & $-11.9{\pm} 10.0$ & $ -0.26{\pm}  0.01$ & 24.4 &33.7 &$ 28.5{\pm} 0.3$ \\
28 & X1.6 & { SOL2014-11-07T16:53} & N17E40 & 12205 & Y & $375.8{\pm} 263.3$ & $ 11.7{\pm}  8.4$ & $309.3{\pm} 19.7$ & $  6.9{\pm}  3.2$ & $  0.4{\pm}  3.4$ & $  4.75{\pm}  0.07$ &  3.6 &35.0 &$  6.0{\pm} 2.7$ \\
29 & M6.1 & { SOL2014-12-04T18:05} & S20W31 & 12222 & N & $237.6{\pm} 183.3$ & $  2.1{\pm}  1.7$ & $ 85.4{\pm}  9.8$ & $ 10.7{\pm}  9.8$ & $ -6.9{\pm} 11.1$ & $  0.05{\pm}  0.01$ &  1.1 &36.9 &$ 11.7{\pm} 0.4$ \\
30 & M6.9 & { SOL2014-12-18T21:41} & S11E10 & 12241 & Y & $243.2{\pm} 162.3$ & $  9.1{\pm}  6.1$ & $374.0{\pm} 16.5$ & $  4.2{\pm}  2.8$ & $  1.3{\pm}  3.6$ & $  2.80{\pm}  0.03$ &  0.1 &38.6 &$  6.0{\pm} 0.2$ \\
31 & X1.8 & { SOL2014-12-20T00:11} & S19W29 & 12242 & Y & $267.4{\pm} 134.4$ & $ 14.8{\pm}  7.6$ & $547.7{\pm} 36.4$ & $  4.8{\pm}  3.7$ & $  1.8{\pm}  4.5$ & $  7.40{\pm}  0.06$ &  3.1 &24.4 &$  3.1{\pm} 0.0$ \\
32 & X2.2 & { SOL2015-03-11T16:11} & S17E22 & 12297 & Y & $326.2{\pm} 144.8$ & $  6.6{\pm}  2.8$ & $207.0{\pm}  9.8$ & $  3.2{\pm}  2.2$ & $  3.2{\pm}  2.2$ & $  5.49{\pm}  0.04$ &  3.0 &17.7 &$ 13.9{\pm} 1.4$ \\
33 & M6.6 & { SOL2015-06-22T17:39} & N13W06 & 12371 & Y & $235.1{\pm} 55.8$ & $ 10.4{\pm}  2.5$ & $443.6{\pm} 15.6$ & $  4.5{\pm}  1.8$ & $  2.1{\pm}  2.2$ & $  7.67{\pm}  0.04$ & 11.6 &24.9 &$  4.6{\pm} 0.0$ \\
34 & X2.2 & { SOL2017-09-06T08:57} & S08W32 & 12673 & N & $481.4{\pm} 236.9$ & $  5.4{\pm}  3.0$ & $105.2{\pm} 11.0$ & $  8.3{\pm}  3.6$ & $  5.8{\pm}  4.9$ & $  6.72{\pm}  0.47$ &  5.1 &12.1 &$  5.9{\pm} 0.3$ \\
35 & X9.3 & { SOL2017-09-06T11:53} & S09W34 & 12673 & Y & $573.4{\pm} 184.3$ & $ 16.6{\pm}  6.7$ & $276.8{\pm} 31.2$ & $  8.5{\pm}  4.8$ & $  2.5{\pm}  5.2$ & $ 16.83{\pm}  0.94$ &  2.6 & 7.3 &$  2.5{\pm} 1.0$ \\
\hline
\hline 
\enddata
\hypertarget{tb1:a1} {$^{a}$}CME association of the flares. ``Y'' (``N'') refers to yes (no). \\ 
\hypertarget{tb1:b1} {$^{b}$}IRS (initial ribbons separation) is averaged in the very early flare stage, with $\sigma$ taken as the error. 
See calculation details for all parameters in {Table~\ref{tb:paras}}. \\ 
\end{deluxetable*}
\end{longrotatetable}%
}

\renewcommand{\arraystretch}{1.1}
\setlength{\tabcolsep}{0.2mm}{


\begin{deluxetable*}{cccccc} 

\tabletypesize{\footnotesize} 


\tablecaption{Parameters quantifying $B_h$-increase and flare ribbons}\label{tb:paras} 
\tablehead{
\colhead{Parameter$^{\hyperlink{tb2:a}{a}}$\phantom{a}} & \colhead{Description\phantom{a}} & \colhead{Calculation$^{\hyperlink{tb2:b}{b}}$\phantom{a}} & \colhead{Unit\phantom{aa}} & \colhead{Type$^{\hyperlink{tb2:c}{c}}$\phantom{a}} & \colhead{Change$^{\hyperlink{tb2:d}{d}}$} 
}
\startdata 
 $\overline{B}_h$ & Mean $B_h$  & $\displaystyle \overline{B}_h{=}\frac{1}{n} \Sigma B_h$ & Gauss & Intensive & $\Delta \overline{B}_h$  \\ 
 $\Sigma B_h$ & Total $B_h$  & $\Sigma B_h{=}{\int} B_h dA$ & Mx & Extensive & $\Delta \Sigma B_h$  \\ 
 Area & Area & $\displaystyle \Sigma dA$ & Mm$^2$ & Extensive & Area  \\
 $\overline{\theta}$ & Mean inclination angle & $\displaystyle \overline{\theta}{=}\frac{1}{n}\Sigma arctan(\frac{B_x}{B_y})$ & Degree & Intensive &  $\Delta \overline{\theta}$  \\  
 $\overline{S}$ & Mean shear angle & $\displaystyle \overline{S}{=}\frac{1}{n}\Sigma arccos(\frac{\mathbf{B_{pot}\bullet B_{obs}}}{|B_{pot}||B_{obs}|})$ & Degree & Intensive &  $\Delta \overline{S}$  \\  
 $\rho_{tot}$ &  \makecell{Proxy for total photospheric\phantom{a}\\magnetic free energy density\phantom{a}} & $\rho_{tot}{=}\displaystyle \frac{1}{8\pi}\Sigma (\mathbf{B_{obs}-B_{pot}})^2 dA$\phantom{a} & ${\rm erg\ cm^{-1}}$ & Extensive &  $\Delta \rho_{tot}$   \\ 
\hline 
 $\tau$ & Timescale of $B_h$-increase & \makecell{Quoted from the step-like function (Equation~\ref{eq:step}) \\fitting of $\overline{B}_h$ evolution} & Minutes & &    \\
 $t_{start}$ & Start time of $B_h$-increase & As above & \makecell{Minutes \\(since flare start)} & &    \\ 
\hline 
 IRS & Initial ribbon separation & \makecell{Ribbons separation averaged over a duration of \\5 minutes before and 1 minute after the flare start} & Mm & &    \\
\hline
\enddata
\hypertarget{tb2:a} {$^{a}$}  
Area is calculated from $B_h$-increase region identified at each moment.  
$\overline{B}_h$, $\Sigma B_h$, $\overline{\theta}$, $\overline{S}$, and $\rho_{tot}$ are computed using the post-flare $B_h$-increase mask.\\  
\hypertarget{tb2:b} {$^{b}$} 
Here $n$ is the number of pixels of $B_h$-increase region.  
$B_{obs}$ is the observed field; $B_{pot}$ is the potential field calculated  
by a Fourier transformation method~\citep{Alissandrakis_1981}. The formulae for $\overline{\theta}$, $\overline{S}$, and $\rho_{tot}$ are adapted from~\citet{Bobra_2014}.\\
\hypertarget{tb2:c} {$^{c}$}``Intensive''-type and ``extensive''-type parameters measure the intensity and extension of corresponding physical quantity, respectively. \\ 
\hypertarget{tb2:d} {$^{d}$} Parameters change throughout flares. $\Delta \overline{B}_h$ and $\Delta \Sigma B_h$ are quoted from corresponding fittings. 
$\Delta \overline{S}$, $\Delta \overline{\theta}$, and $\Delta \rho_{tot}$ are calculated through subtracting the pre-flare value (averaged in five minutes prior to the flare) from the post-flare value (averaged in five minutes after the flare). See relevant details in Section~\ref{subsec:exa}. \\
\end{deluxetable*}
}


\acknowledgments{
We thank Haisheng Ji for the helpful discussions. 
We thank our anonymous referee for his/her constructive comments that significantly improved the manuscript. 
We acknowledge the {\it SDO}, {\it SOHO}, and {\it GOES} missions for providing quality observations. 
L.L. is supported by the Guangdong Basic and Applied Basic Research Foundation. 
}


\bibliographystyle{aasjournal} 

\end{document}